\documentstyle[12pt,epsfig]{article}
\newcommand{\blankline}{\vskip .3cm}
\newcommand{\f}{\begin{equation}}
\newcommand{\ff}{\end{equation}}
\begin{document}
\centerline{\LARGE  The
cubic matrix model }
\blankline
\centerline{\LARGE and a duality} 
\blankline
\centerline{\LARGE between strings and loops}
 
\blankline
\centerline{Lee Smolin}
\blankline
\centerline{\it  Center for Gravitational Physics and Geometry}
\centerline{\it Department of Physics}
\centerline {\it The Pennsylvania State University}
\centerline{\it University Park, PA, USA 16802}
\centerline{and}
\centerline{\it The Blackett Laboratory}
\centerline{\it Imperial College of Science, Technology and Medicine}
\centerline{\it South Kensington, London SW7 2BZ, UK}
\centerline{smolin@phys.psu.edu}
\blankline
\blankline
\centerline{June 6, 2000}
\blankline
\blankline\blankline
\blankline
\centerline{ABSTRACT}
We find evidence for a duality between the standard matrix formulations of
$\cal M$ theory and a background independent theory which
extends loop quantum gravity by replacing $SU(2)$ with a supersymmetric
and quantum group extension of $SU(16)$.  
This is deduced from the recently proposed cubic matrix model
for $\cal M$ theory which has been argued to have compactifications
which reduce to the IKKT and dWHN-BFSS matrix models.  Here we
find new compactifications of this theory whose Hilbert spaces consist  
of $SU(16)$ conformal blocks on compact two-surfaces.  These 
compactifications break the $SU(N)$ symmetry of the standard $\cal M$ theory
compactifications, while preserving $SU(16)$, while the BFSS model 
preserve the $SU(N)$ but break $SU(16)$ to 
the $SO(9)$ symmetry of the 11 dimensional light cone
coordinates.  These results suggest that the supersymmetric and quantum
deformed $SU(16)$ 
extension of loop quantum gravity provides a dual, background
independent description of the degrees of freedom and dynamics of the
$\cal M$ theory matrix models.

\vfill
\eject

\section{Introduction}

One of the oldest and deepest ideas in gauge theories is the
conjectured duality between a loop description, based on the
Wilson loops, or holonomies, of a gauge theory and a string 
description, in which the position of a string of
quantized electric flux are taken as the fundamental 
coordinates\cite{basicdual}.
Taken into the gravitational context, this suggests that there
should be a duality between string theory and
loop quantum gravity, as the latter is based on the quantum
dynamics of the Wilson loops of the spacetime 
connection\cite{lp1}-\cite{loopreview}.
In this paper we propose a specific
form for such a string/loop duality, by finding evidence that a 
particular extension
of loop quantum gravity is  dual to the standard dWHN-BFSS\cite{CH,dWHN,BFSS}
and IKKT\cite{IKKT} matrix models.  We do this by arguing that both
arise from different compactifications of a single matrix
model, called the cubic matrix model\cite{cubic}.

In a recent paper the cubic matrix models were proposed,
and we presented some evidence that one of them has compactifications
which reproduce the IKKT and dWHN-BFSS 
matrix models\cite{cubic}. 
In this paper we 
study  a new class of compactifications of these
models, which lead directly to a background independent description of the
theory. This turns out to be an extension of loop quantum
gravity in which the usual $SU(2)$ algebra has been replaced
by a supersymmetric and quantum extension of $SU(16)$.  The
$SU(16)$ means that the model extends the symmetries of the
$9$ dimensional Clifford algebra, while the quantum
group extension means that it is also in a class of theories
previously proposed as background independent membrane field
theories\cite{tubes,pqtubes}.  In \cite{mpaper} we argued that
when based on an $SU(16)$ symmetry this may provide a  
background independent formulation of $\cal M$ theory.  Here
we provide independent evidence for this claim by showing
how it can be obtained by compactification of a model that
we argue has other compactifications that lead to 
the standard $\cal M$ theory matrix models.
This suggests that the full duality group of $\cal M$ theory
relates the standard background dependent 
descriptions of $\cal M$ theory to this new class of
background independent theories.

The model we defined in \cite{cubic} is based on two simple ideas.
First the degrees of freedom are an
$N\times N$ matrix, $M$, whose elements themselves are matrices,
valued in $Osp(1|32)$.  Thus, the matrix elements refer to elements
of an algebra rather than to positions in a flat background
geometry.  The different background geometries must then arise from 
expansions around classical solutions that break the algebra to one
that generates the symmetry group of a background spacetime.  The
algebra $Osp(1|32)$ is chosen because there is evidence that it may
be the full symmetry group of $\cal M$ theory\cite{m,Osp(1|32)}.
Second, the action has the
simplest non-trivial form possible, which is the trace of the cube
of $M$.  

In \cite{cubic} we argued that this theory has a  compactification
which break the symmetries of $M$ down to the Super-Poincare group in 
$9+1$ dimensions. The one loop effective action describing fluctuations
around the classical solution representing the compactification was
argued to reproduce the $IKKT$ matrix model.  Another compactification
was described, based on a classical solution that breaks the symmetry
to the Super-Euclidean group of the light cone gauge of  
$10+1$ dimensional Minkowski spacetime.  We argued in \cite{cubic}
that the one loop effective action includes the dWHN-BFFS matrix
model\footnote{These claims are supported by one loop calculations
that will be described in detail in \cite{1loop}.}.

The model has, however, many classical solutions in which 
the $Osp(1|32)$  is broken to a group which 
is not the symmetry group of any spacetime.  The small fluctuations
around one of these solutions describes a quantum mechanical system which
cannot be interpreted in terms of a
background spacetime.  We may call these {\it non-geometrical
compactifications}.  As these incorporate more of the 
$Osp(1|32)$ group which has been conjectured to be the full
symmetry group of $\cal M$ theory, it is necessary to understand
these solutions to understand the full physical content of the theory
and the full range of duality transformations which operate on its
solutions. 

Among the nongeometrical compactifications of the cubic matrix model
are a special set
whose continuum limits are  Chern-Simons field theories\cite{cubic}.  
As these
are topological field theories, they are independent of any metric.
Further, quantum Chern-Simons theory is exactly solvable and its 
state space is 
understood in terms of the representation theory of quantum 
groups\cite{ed-cs,ms,verlinde}.
This gives rise to a completely
algebraic description of the
physics of these non-geometrical compactifications.
This can be understood in several different ways, one of which
involves spaces of conformal blocks, or intertwiners of quantum
groups, on 2-surfaces.  As was pointed out first by Crane\cite{louis2d3d},
and Kauffman\cite{lou}, the $k\rightarrow \infty$
limit of this structure (where $k$ is the level, or the coupling
constant of the Chern-Simons theory) reproduces the spin networks which 
label a normalizable bases of diffeomorphism invariant 
states in  loop quantum gravity descriptions of quantum
gravity and supergravity.  We will see here that this can be used
to derive an 
extension of loop quantum gravity from the Chern-Simons compactifications of
the cubic matrix model.

To make these claims precise we must distinguish two meanings of
background independence. We can have a quantum field 
theory that depends on the
topology and differential structure of a manifold of some given
dimension, but is independent of any classical fields.  Such theories
must then have active diffeomorphisms as part of their gauge group.
As a result all fields are represented by quantum operators
which are subject to evolution by dynamical laws.
We may call these {\it field-background independent} theories.  Loop
quantum gravity\cite{lp1,lp2,lp3,loopreview} has provided examples of such 
theories, in several
different dimensions, with and without supersymmetry\cite{slp1,slp2}.

However, one can require that the theory not depend on even the
dimension, topology or differential structure of a manifold. We
may call such a theory {\it manifold-background independent}. A theory
with both properties may be called {\it totally-background independent}.

If string theory has a background independent formulation it must
be totally background independent, because its different solutions
are defined on manifolds of different dimension and topology.  
Loop quantum gravity, in the form originally given in
\cite{lp1,lp2,lp3,loopreview}, is not  such a theory. This is because
its derivation through a rigorous quantization of general relativity
and supergravity required that the manifold, dimension and 
differential structure be fixed.  However,
it is possible to extend the structure of loop quantum gravity to
make it totally background independent\cite{FM2}. As explained in 
\cite{tubes} this requires thickening
the spin networks to 2-surfaces labeled by conformal
blocks, or intertwiners of a quantum group. 
This extension is also motivated by the fact that it solves
a key problem in loop quantum gravity, because it introduces
certain terms in the Hamiltonian or action which are required
both for the existence of a classical limit\cite{trouble1,trouble2} 
and the recovery of spacetime diffeomorphism 
invariance\cite{FM2,CM1}.

It was noted in
\cite{tubes,pqtubes} that the resulting theories may be understood
as  background independent
membrane field theories, and that the $SL(2,Z)$ duality group
can be easily represented in a background independent fashion.  
This recalls the old arguments that the fundamental 
objects in $\cal M$ theory should be membranes\cite{m}.  To realize
this idea in the present context we must 
invent a theory in which the embeddings of the
membrane in different background can be extracted from the
intertwiners of a quantum group, by symmetry breaking.  A
detailed conjecture for how to do this was proposed in \cite{mpaper}.
It was found there that to match the dWHN-BFSS matrix model the
quantum group must be a quantum deformation of a super-Lie
algebra that extends $SU(16)$.
Here we will derive a closely related  theory from a non-geometrical
compactification of the
cubic matrix model, but with a simpler dynamics, which is also
very reminiscent of the dynamics generated by the Hamiltonian
constraint in quantum gravity and supergravity\cite{ham1}. 

The version of the cubic matrix model we study here was introduced
in \cite{cubic}, and is characterized by the fact that it uses
a complexification of the $Osp(1|32)$ degrees of
freedom which is represented in terms of $SU(16,16|1)$ matrices.
We posit here the  simplest possible action,
given by 
\f
I={k \over 4\pi}  TrM^{3}
\ff
where $M_{iA}^{jB}$ is a double matrix with $i,j=1,\ldots,N$
and $A,B=1,\ldots,33$. For fixed $i,j$, $M_{A}^{B}$ is an 
element in the $33$ dimensional adjoint representation of
$SU(16,16|1)$.   This was called the gauged action in
\cite{cubic}.

It is not difficult to carry out the same steps as described
in \cite{cubic} to show that this model has compactifications
that reduce at the one loop level to the IKKT and BFSS models.
This will be discussed very briefly in section 8 below and 
described in full detail elsewhere\cite{1loop}.  We may note
that these compactifications involve breaking the $SU(16,16|1)$
symmetry to $SO(9,1)$ and $SO(9)$, respectively.   
The non-geometric compactifications we will study break
the symmetry in a different way. 
\f
SU(16,16) \rightarrow Sp(2) \oplus SU(16)
\ff

In the next section we introduce the model and some of
its properties.  In section 3 we describe a Chern-Simons
compactification leading to an $SU(16)$ Chern-Simons theory.
In section 4 we show how related compactifications give rise
to a set of interacting Chern-Simons theories.
In sections 5 to 7  we describe the quantization of these
compactifications and show how they reproduce the extension
of loop quantum gravity described in \cite{tubes,pqtubes,mpaper}.
An argument for how the
dWHN-BFFS theory may be recovered from the effective action 
in  a different limit of the theory is sketched in section 8.
This allows us, in section 9 to describe the explicit duality 
that holds between the standard background dependent matrix models
of $\cal M$ theory and the background independent description
derived here.  This allows us to answer questions such as what corresponds
to $D0$ branes in the background independent language.
The paper closes with a brief mention of some of the important
open problems raised by the results reported here.

\section{The $SU(16,16|1)$ matrix theory}

We begin by describing the model and its basic properties.
We recall first the definition of $Osp(1|32)$, which consists
of supermatrices of $32$ even dimensions and one odd dimension,
$M_{A}^{B}$ that preserve the graded antisymmetric metric
\f
G_{A}^{\ B} = 
\left [
\begin{array}{ccc}
    0 & -I & 0  \\
    I & 0 & 0  \\
    0 & 0 & 1
\end{array}
\right ]
\ff
(where the first two rows and columns are $16\times 16$ bosonic 
blocks, while the third row and column is in the one fermionic 
coordinate) so that
\f
M \cdot G = - G \cdot M^{T} 
\ff
where $T$ stands for transpose.
We complexify this by considering complex matrices of the
same form, which satisfy
\f
M \cdot G = - G \cdot M^{\dagger}  
\ff
where $\dagger$ means hermitian conjugate. These generate a supergroup by
$R=exp M $,  
which satisfies $R \cdot G \cdot R^{\dagger}= G$.  This group is
$SU(16,16|1)$, as can be seen from
the fact that $\imath G_{A}^{B}$ is an hermitian metric of
signature $(16,16)$.  We may decompose $M_{A}^{B}$ as
follows, 
\f
M_{A}^{\ B} = 
\left (
\begin{array}{ccc}
    A_{2}+Y & A_{-} & \Psi  \\
    A_{+} & -A_{2}+Y & \Phi  \\
    \Phi^{\dagger} & -\Psi^{\dagger} & 0
\end{array}
\right )
\label{param1}
\ff
where the $A_{2},A_{\pm}$ are three
$16 \times 16$ hermitian matrices, $Y$ is a tracefree $16 \times 16$
antihermitian matrix and $\Psi^{A}$ and
$\Phi^{A^{\prime}}$ are $16$ component spinors. 
We will also find it convenient to use 
$A_{0}=A_{+}-A_{-}$ and $A_{1}=A_{+}+A_{-}$.
The components of $Y$ and $A_{a}$ for $a=0,1,2$  are even Grassman
variables, while the components of $\Phi$ and $\Psi$ are odd
Grassman.  We will decompose the $33$ dimensional indices
$A,B,\ldots$ as $A=(P,P^{\prime},\cdot )$ where
$P=1,\ldots,16,$, $P'=1',\ldots,16',$ and $\cdot$ is the
lone fermionic index.  We will find it very useful to consider
the decomposition  
\f
SU(16,16) \rightarrow Sp(2) \oplus SU(16)
\ff
where the $SU(16)$ is generated by
\f
W_{A}^{\ B} = 
\left (
\begin{array}{ccc}
    Y & 0 & 0  \\
  0 & Y & 0  \\
   0 & 0 & 0
\end{array}
\right )
\label{SU16}
\ff
and the generators of $Sp(2)$ are given by 
\f
\tau^{0} = 
\left (
\begin{array}{ccc}
   0 & -I & 0 \\
    I & 0 & 0  \\
    0 & 0 & 0
\end{array}
\right ) , \ \ \ 
\tau^{1} = 
\left (
\begin{array}{ccc}
   0 & I & 0 \\
    I & 0 & 0  \\
    0 & 0 & 0
\end{array}
\right ) , \ \ \ 
\tau^{2} = 
\left (
\begin{array}{ccc}
   I &0 & 0 \\
    0 & -I & 0  \\
    0 & 0 & 0
\end{array}
\right )
\label{taus}
\ff
These satisfy
\f
{1\over 32} Tr\tau^{a}\tau^{b}= \eta^{ab}
\ff
where $\eta^{ab}$ is the $2+1$ dimensional metric
\f
\eta^{ab} = 
\left (
\begin{array}{ccc}
   -1 &0 & 0 \\
    0 & 1 & 0  \\
    0 & 0 & 1
\end{array}
\right )
\label{eta}
\ff
thus showing $Sp(2)=SO(2,1)$.

The $SU(16)$ transformations can be extended to 
the superalgebra $SU(16|1)$, leading to a reduction of
superalgebras,
\f
SU(16,16|1) \rightarrow Sp(2) \oplus SU(16|1).
\ff

We now extend the matrix by considering each entry to be an
$N\times N$ matrix parameterized by $i,j=1,\ldots,N$.
We thus have a double matrix
$M_{Ai}^{\ Bj}$.  We define the action
\f
I^{gauged}= {k \over 4\pi } Tr M^{3}
\label{gaugedaction}
\ff
where the multiplication and trace is on both sets of indices.
$k$ is the coupling constant of the theory.
This action is somewhat simpler than that studied in \cite{cubic}
and has more gauge symmetry, as we can see by writing it more
explicitly
\f
I^{gauged}= {k \over 4\pi } \sum_{ijk}Tr_{SU(16,16|1)} M_{i}^{\ j} 
\cdot M_{j}^{\ k}
\cdot M_{k}^{\ i}
\label{gaugedaction2}
\ff
where for each $i,j,k$ the multiplications and trace are over the
$SU(16,16|1)$ indices.  The action is then invariant under 
transformations
\f
M_{i}^{\ j} \rightarrow U(i) \cdot M_{i}^{\ j} \cdot U(j)^{\dagger}
\label{gauge16}
\ff
where, for each $i$, $U(i) \in SU(16,16|1)$.  
Similarly, for each $A$ and $B$ we
have the gauge symmetry, suppressing now the $SU(N)$ indices
\f
M_{A}^{\ B} \rightarrow V(A) \cdot M_{A}^{\ B} \cdot V(B)^{\dagger}
\ff
\label{gaugeN}
where $V(A)\in SU(N)$.  For this reason
we call it a gauged matrix action.

It is useful to decompose the action in terms of the 
variables defined in (\ref{param1}). We have
\begin{eqnarray}
I^{gauged}&=& {k \over 4\pi } 
Tr_{ij}Tr_{PQ} \left \{ \epsilon^{abc} ( A_{a}A_{b}A_{c} )
+ YA_{a}A_{b} \eta^{ab} + Y^{3} \right \}  \nonumber \\
&&+ 3 Tr_{ij} \left \{  
\Phi^{P} A_{- P}^{\ Q}\Phi_{Q}- \Psi^{P}A_{+ P}^{\ Q}\Psi_{Q}-
A_{2 PQ} \{ \Psi^{Q}, \Phi^{P}\} + Y_{ PQ} [ \Psi^{Q}, \Phi^{P}]
\right \}
\label{action2}
\end{eqnarray}

We see explicitly here the decomposition into $Sp(2) \oplus SU(16|1)$.

It is also interesting to note that when the matrix elements
are restricted to be real, so that the symmetry is reduced
from $SU(16,16|1)$ to $Osp(1|32)$ the action has a translation
invariance given by 
\f
A_{aPi}^{\ Qj}\rightarrow A_{aPi}^{\prime \ Qj} =
A_{aPi}^{\ Qj} +\delta_{i}^{j} V_{aP}^{\ Q}
\label{translations}
\ff
with $V_{a PQ}=V_{a PQ}^{\dagger}$. 
This is reduced from the symmetry of the model studied in
\cite{cubic} but the remaining translation symmetry includes
that of the dWHN-BFSS and IKKT models, when we identify the
fields of those models with components of $M_{A}^{B}$ as
described in \cite{cubic}.

It is also of interest to consider the theory without the
$Y$ degrees of freedom.  This theory is invariant under
the $Sp(2) \oplus SU(16|1)$ subalgebra of $SU(16,16|1)$. 
Its action is given by

\f
I^{MCS}=I^{gauged}_{Y=0} =
{k \over 4\pi } Tr_{ij} \left \{ \epsilon^{abc} Tr_{PQ} ( A_{a}A_{b}A_{c} )
+ \chi^{\dagger P}\cdot \tau^{a} \cdot A_{a P}^{Q}\chi_{Q}
\right \}  
\label{MCS}
\ff   
We will call this {\it matrix Chern-Simons theory}
in the following.

\section{A simple Chern-Simons compactification}

The Chern-Simons compactification comes about because
for each $a$, $A_{a} = (A_{aP}^{\ Q})_{i}^{\ j}$ is an
$N\times N$ matrix of $16$ dimensional hermitian matrices.
There are then compactifications of the model in which 
the $\epsilon^{abc} ( A_{a}A_{b}A_{c} )$ term in the
action becomes an $SU(16)$ Chern-Simons theory. To see this
we compactify the theory on a three-torus, making use of a modification of 
the route studied in \cite{cubic}.  
We consider the classical solutions given by
\f
A_{aiP}^{\ jQ}= \delta_{P}^{Q} ( \partial_{a})_{i}^{j}
\ff
\f
Y= 0
\ff
with the fermionic fields vanishing.
We divide the indices so as to make three derivative
operators. We choose $i=i_{0},i_{1},i_{2}$ where
$i_{a}=-M_{a},\ldots, 0, 1 , \ldots , M_{a}$
such that $N=\prod_{a=0,1,2}(2M_{a}+1)$.  We then choose
\f
(\partial_{0})_{i}^{j} = (\partial_{0})_{i_{0}i_{1}i_{2}}^{j_{0}j_{1}j_{2}}
=i_{0}\delta_{i_{0}}^{j_{0}}\delta_{i_{1}}^{j_{1}}\delta_{i_{2}}^{j_{2}}
\ff
and similarly for the other two derivative operators.
Clearly we 
have $[\partial_{a}, \partial_{b}] =0$. We then expand
around this classical solution, using the usual matrix
compactification trick\cite{trick,BFSS}, defining variables
\f
A_{aiP}^{\ jQ}= \delta_{P}^{Q}( \partial_{a})_{i}^{j} + a_{aiP}^{\ j Q}
\ff
Following the usual translation into continuum fields\cite{trick,BFSS}, 
we find in the limit $M_{a}\rightarrow \infty$
\begin{eqnarray}
I^{gauged}&=& {k \over 4\pi } 
\int_{T^{3}}  Tr_{SU(16)} 
\left \{  
a \wedge da + 
{2\over 3} a^{3}+  \chi^{\dagger}\tau^{a} {\cal D}_{a} \chi
\right \} 
\nonumber \\
&&+(Y_{P}^{Q})^{3}+ Y_{PQ} \eta^{ab} ({\cal D}_{a} \tilde{a}_{b})^{QP} +
Y_{PQ}[\Psi^{P},\Phi^{Q} ] 
\label{csaction}
\end{eqnarray}
where $\tilde{a}_{b}^{PQ}$ is tracefree.  

If we neglect the coupling to the $Y$ field this is a 
Chern-Simons theory on $T^{3}$. 

Note that had we begun with the bosonic part matrix Chern-Simons theory
(\ref{MCS}), whose action is just $\epsilon^{abc}Tr(A_{a}A_{b}A_{c})$, 
the result would just be the first two terms of
(\ref{csaction}).  In this case something remarkable has
happened, which is that the dependence on a background metric
on the torus, which is implicit in the definition of the
compactification, has gone away in the limit $M_{a}\rightarrow \infty$,
leaving a topological field theory.  We may note that with
the ultraviolet cutoff $l_{Planck}$ held fixed, this is equivalent
to the limit in which all three compactification radii are taken
to infinity.  The fact that a topological field theory
emerges from the limit is a consequence of the large amount
of gauge symmetry in the original action, together with the
taking of a limit in which the length scales set by the 
compactification radii are removed. 

The coupling to the spinor variables $\chi$ depends on the 
background $\tau^{a}$ which define the compactification directions.
These result in a supersymmetrization of the state space of
Chern-Simons theory as we will discuss in \cite{superthis}.

The theory with the  $Y$ terms present is more subtle, but
it is possible that they will not change the topological character
of the theory in the limit of infinite compactification radii, up to
renormalizations of the coupling constant of Chern-Simons theory.
The reason is that we can use the gauge symmetry (\ref{gauge16}) to set
$Y_{i}^{j}=0$ {\it locally} in $i$ and $j$. Given any sequence
of values for $i$, with no repeats, 
given by $i_{\mu}=i_{1},\ldots, i_{L}$, we can set
$Y_{i_{\mu}}^{i_{\mu+1}}=0$.  To see this note that by (\ref{gauge16})
\f
\delta Y_{iP}^{jQ}= \delta M_{i \alpha P}^{j \alpha Q}=
U(i)_{\alpha P}^{\gamma R} M_{i \gamma  R}^{j \alpha Q}
-M_{i \alpha P}^{j \gamma R } U(j)_{\gamma R}^{\alpha Q}
\ff
Thus, $Y_{i}^{j}$ transforms like a gauge potential, it can
be set to zero along any open curve in the space labeled
by $i,j$ but its effect on closed curves in those indices 
must be taken into account.
In a matrix compactification the indices $i$ and $j$ refer to 
Fourier modes, in particular the local field is constructed
from sums such as 
$Y(x)= \sum_{k} Y_{i_{0}}^{i_{o}+k} e^{-\imath k\cdot x }$.   
This means we have
a kind of gauge invariance in momentum space. As a result, we can eliminate
the $Y$'s from any scattering matrix involving other degrees of
freedom.  But we cannot eliminate closed loops of $Y$'s in
any Feynman diagram, as they will correspond to a closed loop
of matrix entries. The effect on the classical equations of
motion will only be through the $Y$ equation itself. The 
bosonic equations of motion which follow from (\ref{csaction})
have the form 
\f
\epsilon^{abc}F_{bc}-\chi^{\dagger}\tau^{c}\chi -\eta^{cd}
{\cal D}_{d}Y 
=0
\ff
\f
Y^{2}_{PQ}+ \left ( {\cal D}_{a} a_{b}
\right )_{PQ} \eta^{ab} + [\Psi_P,\Phi_Q ] =0
\ff
If we set $Y=0$ at a point we have the equations of 
Chern-Simons theory plus the condition
\f
\left ( {\cal D}_{a} a_{b}
\right )_{PQ} \eta^{ab} + [\Psi_P,\Phi_Q ] =0
\ff
But this differs only by  the fermion term from the standard 
gauge fixing term which is used to define Chern-Simons theory
perturbatively.  This means that at least up to the effects of
closed loops in the $Y$ variables, the theory given by 
(\ref{csaction}) is a supersymmetric
extension of the $SU(16)$ Chern-Simons theory.   

This does not completely resolve the question of the influence of
the $Y$ degrees of freedom.  However, it suggests that 
the effect of the $Y$'s on the Chern-Simons compactification
can only be to renormalize the coupling of the 
Chern-Simons theory.  For this reason we will neglect the
effects of the $Y$'s in the next sections where we study
the physics of the Chern-Simons compactifications.

\section{The $(\mbox{Chern-Simons})^{P}$ compactification}

We now introduce a different set of compactifications that
reduce the theory to a coupled set of $P$ Chern-Simons theories,
for any $P$. 
The idea is to blow up each of the $i_a-j_a$ entries of the
previous compactification into $P \times P$ blocks.  We then
have a symmetry, for each $i_a,j_a$, which is 
$Sp(2) \oplus SU(16) \oplus U(P)$.  
When we make a compactification on the three torus, as
just described, we find a $3$ dimensional quantum field theory
with symmetry $SU(16) \oplus U(P)$.  There are however,
several different ways of taking the limit that defines
the field theory, which result in a different set of
fluctuating fields with different symmetry.  One way,
which we will sketch in section 8, preserves the $U(P)$
symmetry and breaks the $SU(16)$ down to 
the $SO(9)$ symmetry of the lightcone
gauge of $10+1$ dimensional spacetime. 

Here we will consider a way to preserve the full
$SU(16)$, but break the $U(P)$ symmetry completely.
This leads not to one $3d$ quantum field theory with
$U(P)$ symmetry, but to $P$ 
$3d$ field theories, each of which defines a Chern-Simons
theory.  These interact via bi-local fields that create and
annihilate pairs of punctures that join the tori on which
these Chern-Simons theories are defined.
As we will show in the next several sections,
this gives a background independent theory. This theory is
a version
of the background independent membrane  
field theory\cite{tubes,pqtubes} which was proposed in
\cite{mpaper} as a background independent version of
$\cal M$ theory.

We begin the demonstration by choosing integer factors  
$M_{a}$ and $P$ such that 
$\prod_a (2M_a +1) P=N$.     
We will write
\f
i= i_{0},i_{1},i_{2},I
\ff
with $I =1,\ldots,P$ and $i_{a}=- M_a,...,M_a $.  

We then decompose $M_{I}^{\ J}$ according to
(with all the other indices suppressed)
\f
M_{I}^{\ J} = 
\left [
\begin{array}{cccc}
    A^{1} &  B_{1}^{2} &  B_1^3 & ... \\
    B_2^1  & A^2    &    B_2^3 & ...  \\
    ...   & ... & A^3       & ...
\end{array}
\right ]
\ff
That is, we define
\f
A_{a I i_{0}i_{1}i_{2} P}^{ \ J=I j_{0}j_{1} j_{2} Q} = 
A_{a  i_{0}i_{1}i_{2} P}^{ \ I j_{0}j_{1} j_{2} Q}
\ff
and for the off diagonal terms,
\f
A_{a I i_{0}i_{1}i_{2} P}^{ \ J \neq I j_{0}j_{1} j_{2} Q} 
\tau^{a\beta}_{\alpha}= 
B_{\alpha I  i_{0}i_{1}i_{2} P}^{ \ \beta J j_{0}j_{1} j_{2} Q}
\ff

We continue to neglect the role of the  $Y$ field..
The dynamics is given by 
(with the $i_{a}$ indices suppressed)
\begin{eqnarray}
        I^{gauged}&= & {k \over 4\pi } Tr_{i_{a}} \left \{ 
    \sum_{I} 
\epsilon^{{abc}}(A_{a P}^{I \ Q} A_{b Q}^{I \ R}  A_{c R}^{I \  P}  )
+ \chi^{\dagger P }_{\alpha I }\tau^{a \alpha}_{\beta} A_{a P}^{\ I Q}
\chi^{\beta I }_{Q}
\right. \nonumber \\
&& + \sum_{J\neq I}\left [  3 B_{ I P \alpha }^{J Q \beta} 
\tau^{b \gamma}_{\beta} A_{b Q}^{J \ R} B_{ J R \gamma }^{I P  \alpha } 
\right. \nonumber \\ 
&&\left. + \chi^{\dagger P J}_{\alpha I}\tau^{a \alpha}_{\beta} A_{a P}^{\ J Q}
\chi^{\beta I}_{ Q J}
+ 
\chi^{\dagger P J}_{\alpha }B_{ P J}^{\ I Q}
\chi^{\beta J}_{ Q I}
+
\chi^{\dagger P J}_{\alpha I}B_{ P J}^{\ I Q}
\chi^{\beta I}_{ Q } \right ]
\nonumber \\
&& + \left. \sum_{J\neq I\neq K } \left [ B_{ I P \alpha }^{J Q \beta} 
B_{J Q \beta }^{K  R \gamma } B_{ K R \gamma }^{I P  \alpha } 
+
\chi^{\dagger P J}_{\alpha I}B_{ P J}^{\ K Q}
\chi^{\beta I}_{ Q K} \right ]
\right \} 
\end{eqnarray}    
We see we get $P$ theories described by an $A^{I}$ coupled by terms
involving the $B_{I}^{J}$ and $\chi^{\beta I}_{ Q K}$ variables. 

The theory has a  $U(P)$ symmetry mixing up the $I,J$ indices.
We now define a class of compactifications which can lead to 
a quantum theory that breaks this symmetry.
To do this we will define a Chern-Simons compactification to an
$SU(16)$ Chern-Simons theory separately for each $I$.  For each
of the $P$ diagonal
 $A^{I}$'s we define 
the compactification to an $SU(16)$ gauge theory 
described in the last section. This leads,
for each $I$, to 
a Chern-Simons theory on a three torus based on the 
fusion algebra
of $SU(16) $. 

This multiple-Chern-Simons compactification also induces
a transformation on the $B$ variables.  
We find (for $J\neq I$) that when $M_{a} \rightarrow \infty$
the $ B_{I \alpha_{I}i}^{J \beta_{J} j}$ become bilocal 
fields whose domain are  pairs of the $T^{3}$ on which the
Chern-Simons theories are defined.
\f
 B_{I \alpha_{I}i_{0}i_{1}i_{2}P}^{J \beta_{J} j_{0}j_{1}j_{2}Q} \rightarrow  
 B_{I \alpha_{I}P}^{J \beta_{J} Q} (x_{I}, x_{J})
\ff
We also have bilocal fermionic fields given by
\f
\chi_{Q I i_{0}i_{1}i_{2}}^{\alpha J j_{0}j_{1}j_{2}}
\rightarrow \chi_{Q I}^{\alpha J } (x_{I}, x_{J})
\ff

Thus after the $P$ simultaneous three-torus compactifications
and the $M_{a} \rightarrow \infty $ limits the action of the theory becomes
\begin{eqnarray}
I^{cgauged}&=&  {k \over 4\pi } \sum_{I} S_{CS}^{I}({\cal 
A}^{I},\chi^{I})    
+ \sum_{J\neq I} {k \over 4\pi } \int_{T^{3}_{I}}d^{3}x_{I}
\int_{T^{3}_{J}}d^{3}x_{J} \left \{   Tr_{G}\left (  
B_{I \alpha}^{J \beta } (x_{I}, x_{J})
 \tau^{\bar{a} \gamma}_{\beta} {\cal D}_{\bar{a}}^{J}  
B_{J \gamma}^{I \alpha } (x_{J}, x_{I})  \right )  \right. \nonumber \\
&&+ \chi^{\dagger P J}_{\alpha I} (x_{I},x_{J}) \tau^{a \alpha}_{\beta} {\cal D}_{a P}^{\ J Q}
\chi^{\beta I}_{ Q J} (x_{J},x_{I})
+ 
\chi^{\dagger P J}_{\alpha }(x_{J)} B_{ P J}^{\ I Q} (x_{J},x_{I})
\chi^{\beta J}_{ Q I} (x_{I},x_{J})
\nonumber \\
&&
\left. +
\chi^{\dagger P J}_{\alpha I} (x_{I},x_{J}) B_{ P J}^{\ I Q} 
(x_{J},x_{I})
\chi^{\beta I}_{ Q } (x_{I}) \right \}
\nonumber  \\
&&+ \sum_{I\neq J \neq K} {k \over 4\pi } \int_{T^{3}_{I}}d^{3}x_{I}
\int_{T^{3}_{J}}d^{3}x_{J}  \int_{T^{3}_{K}}d^{3}x_{K} Tr_{G}\left (  
B_{I \alpha}^{J \beta } (x_{I}, x_{J}) 
B_{J \beta}^{K \gamma } (x_{J}, x_{K})
B_{K \gamma}^{I \alpha } (x_{K}, x_{I}) \right.
\nonumber \\
&&+ \left.
\chi^{\dagger P J}_{\alpha I}(x_{I},x_{J})B_{ P J}^{\ K Q}(x_{J},x_{K})
\chi^{\beta I}_{ Q K}(x_{K},x_{I})
\right ) 
\label{CSP}
\end{eqnarray}     
where
\f
S_{CS}^{I}({\cal A}^{I},\chi^{I}) = 
\int_{T^{3}}  Tr_{SU(16)} \left \{  a \wedge da + 
{2\over 3} a^{3}+  \chi^{\dagger}\tau^{a} {\cal D}_{a} \chi
\right \} 
\ff
Thus, we have $P$ Chern-Simons theories, each defined on a distinct 
3-torus, which 
interact with each other via the bi-local fields 
$B_{I \alpha}^{J \beta } (x_{I}, x_{J})$.

\section{Hamiltonian dynamics of the  $(\mbox{Chern-Simons})^{P}$
compactification}

We now study the dynamics of the theory defined by the 
multiple Chern-Simons compactification. To simplify the
discussion we will ignore the fermion fields and
consider only the bosonic parts of (\ref{CSP}).  The fermion
terms give a supersymmetric completion of the structure we will
describe here; details of this will be described 
elsewhere\cite{superthis}.

To uncover the dynamics of the theory in this compactification
we make a $2+1$ splitting in each of the $P$ 3-tori. This
gives coordinates $x^{{a}}_{I}= (t_{I}, x_{I}^{m})$,
where we define spatial indices $l,m,n=1,2$.  
We also let the compactification radius for each time coordinate
go to infinity, so that the Hamiltonian theory becomes defined
on a domain which is $P$ copies of $R \times T^{2}$.
The action
then takes the form,
\begin{eqnarray}
I^{cgauged}&=& \sum_{I}{k \over 4\pi } \int dt_{I} \int d^{2}x_{I} 
Tr_{SU(16)} \left [
\epsilon^{mn} 
{A}^{I}_{m} \dot{A}^{I}_{n} -{ A}^{I}_{0} {\cal G}^{I}
\right ]   \\
&+&  \sum_{ J > I} {k \over 4\pi } \int dt_{I}¥   d^{2}x_{I} 
 dt_{J}  d^{2}x_{J}Tr \left \{
(B_{J}^{I  } (x_{J},t_{J};  x_{I},t_{I}))^{\ \beta}_{\alpha}
 \tau^{0 \gamma}_{\beta} 
 \left ( {\partial \over \partial t_{I}}  - {\partial \over \partial  t_{J}} 
 \right ) 
B_{I \gamma}^{J \alpha } (x_{I}, t_{I}, x_{J}, t_{J}) \right. \nonumber \\
 &+&   \left.
(B_{J}^{I  } (x_{J},t_{J};  x_{I},t_{I}))^{\ \beta}_{\alpha}
 \tau^{m \gamma}_{\beta}  
 \left ( {\cal D}_{m}^{I}- {\cal D}_{m}^{J} \right )
B_{I \gamma}^{J \alpha } (x_{I}, t, x_{J}, s) ]
\right \} 
\nonumber \\
&+& \sum_{I\neq J \neq K} {k \over 4\pi } \int_{T^{3}_{I}}d^{3}x_{I}
\int_{T^{3}_{J}}d^{3}x_{J}  \int_{T^{3}_{K}}d^{3}x_{K} Tr \left (  
B_{I \alpha}^{J \beta } (x_{I}, x_{J}) 
B_{J \beta}^{K \gamma } (x_{J}, x_{K})
B_{K \gamma}^{I \alpha } (x_{K}, x_{I})
\right ) \nonumber
\label{2+1form}
\end{eqnarray}
where the Gauss's law constraint is
\begin{eqnarray}
{\cal G}^{I}(x_{I},t)&=& {1\over 2}\epsilon^{mn} F_{mn}^{I}(x_{I},t) 
+\sum_{J> I}\int dt_{J}
\int d^{2}x_{J} 
(B_{J}^{I  } (x_{J},t_{J};  x_{I},t_{I}))^{\ \beta}_{\alpha}
\tau^{0 \gamma}_{\beta}   
B_{I \gamma}^{J \alpha } (x_{I}, t_{I}, x_{J}, t_{J}) \nonumber \\
&- &
\sum_{K< I}\int dt_{K}
\int d^{2}x_{J} 
(B_{I}^{K  } (x_{I},t_{I};  x_{K},t_{K}))^{\ \beta}_{\alpha}
\tau^{0 \gamma}_{\beta}   
B_{K \gamma}^{I \alpha } (x_{K}, t_{K}, x_{I}, t_{I})
\end{eqnarray}
We see that the theory appears to be non-local in time.  We will
shortly see that this is an expression of a many fingered
time gauge invariance, in that the time can be evolved independently
on each torus.

The $A^{I}_{m}$ have conventional momenta,
\f
\pi^{Im}(x_{i},t) \equiv {\delta I^{cubic}\over \dot {\cal 
A}^{I}_{m}(x_{I},t)} = {k \over 2\pi }\epsilon^{mn}{ A}_{n}^{I}(x_{I},t)
\ff
The $B_{I}^{J}$'s depend on two time coordinates, one in 
each of the two tori they live in.  The momenta then similarly
depend on two time and two space variables. As we see from 
(\ref{2+1form}) the momenta of the $B_{I}^{J}$'s depends on the
difference of its two time coordinates,
\f
\Pi_{\ J}^{I}(x_{I},t_{I};x_{J},t_{J})_{\alpha P}^{\ \gamma Q} \equiv
{\delta I^{cubic}\over (\partial t_{I}-\partial t_{J} ) 
{B}_{I \gamma Q}^{J \alpha P} (x_{I}, t_{I}, 
x_{J}, t_{J}) }
= {k \over 4\pi }
({B}_{J}^{I  })^{ \beta Q}_{\alpha P} (x_{J}, t_{J}, x_{I}, t_{I}) 
\tau^{0 \gamma}_{\beta}
\label{Bmomenta}
\ff

To define a polarization we 
pick an arbitrary ordering of the $P$ tori.  This breaks the
$U(P)$ symmetry.
We then take the variables ${B}_{I \gamma}^{J \alpha } (x_{I}, t, x_{J}, s) $
for $J>I$ to be configuration variables, while the momenta are
coded in the $B_{I}^{J}$ for $J<I$.

Note also that there is some freedom of the density transformation properties
of the fields  defined by the compactification.  This must
be defined so that all the integrands are densities.  A sensible
choice seems to be to define the compactification in such a way 
that the $B_{I \gamma}^{J \alpha } 
(x_{I}^{{a}}, x_{J}^{{b}} )$ are densities on
the second spacetime variables $ x_{J}^{{b}}$ and ordinary
functions on the first. The momenta will then have the opposite
density weights, i.e. weight one in the first slot and zero in the
second.  

Finally, we write the action in Hamiltonian form
\begin{eqnarray}
I^{gauged}&=& \sum_{I}  \int dt_{I} \int d^{2}x_{I} Tr_{SU(16)} \left [
\pi^{Im} \dot{ A}^{I}_{m} -{ A}^{I}_{0} {\cal G}^{I}
\right ]   \\
&&+  \sum_{ J > I}  \int dt_{I}   d^{2}x_{I} 
 dt_{J}  d^{2}x_{J}Tr_{SU(16)} \left 
 [ \Pi_{I}^{\ J}(x_{I},t_{I};x_{J},t_{J})_{\alpha}^{\beta}
  \left ( {\partial \over \partial t_{I}}  - {\partial \over \partial 
 t_{J}} \right )
 {B}_{I \gamma}^{J \alpha } (x_{I}, t_{I}, x_{J}, t_{J}) \right ]
\nonumber \\
&&+ \sum_{ J>I}  \int dt_{I} dt_{J} {\cal H}^{IJ}_{2}(t_{I},t_{J})
+\sum_{I<J<K}  \int dt_{I} dt_{J} dt_{K}{\cal 
H}^{IJK}_{3}(t_{I},t_{J},t_{K})
\end{eqnarray}    
where the Gauss's law constraint is now
\f
{\cal G}^{I}(x_{I},t) = {1\over 2}\epsilon^{mn} F_{mn}^{I}(x_{I},t) 
- j^{I}(x_{I},t)=0
\label{gauss}
\ff
where the $SU(16)$ current for each Chern-Simons theory is given by 
\f
j^{K P}_{Q}(x_{K},t_{K})= 
\sum_{I<K}\int d^{3}x_{I}Tr_{Sp(2)}\Pi_{IR}^{KP}\cdot B_{IQ}^{KR}-
\sum_{K<J}\int d^{3}x_{J}Tr_{Sp(2)}\Pi_{KR}^{JP}\cdot B_{KQ}^{JR}
\label{current}
\ff
and the two and three time Hamiltonians are given by
\f
{\cal H}^{IJ}_{2}(t_{I},t_{J})= \int d^{2}x_{I}d^{2}x_{J} Tr_{SU(16)}
\bar{\Pi}_{I}^{\ J}(x_{I},t;x_{J},s)_{\alpha}^{\beta}  
\tau^{m \gamma}_{\beta}  
 \left ( {\cal D}_{m}^{I}- {\cal D}_{m}^{J} \right )
B_{I \gamma}^{J \alpha } (x_{I}, t, x_{J}, s) 
\label{H2}
\ff
\f
{\cal H}^{IJK}_{3}(t_{I},t_{J},t_{K})= 
3 \int d^{2}x_{I}d^{2}x_{J}d^{2}x_{K}  Tr 
\left \{ B_{I \gamma}^{K \alpha }
\bar{\Pi}_{J\beta }^{\ K \gamma} 
\bar{\Pi}_{I\gamma }^{\ J \alpha}   + B_{I \alpha}^{J \beta } 
B_{J \beta}^{K \gamma } 
\bar{\Pi}_{I\gamma }^{\ K \alpha }
\right \}
\label{H3}
\ff
Here $\bar{\Pi}_{I \alpha }^{\ J \beta} =
\Pi_{I \alpha }^{\ J \gamma } \tau^{0 \beta}_{\gamma}$.

\section{Quantization of the $(CS)^{P}$ compactification}

What we have is a multi-time theory in which a set of $P$
Chern-Simons theories, each with a time coordinate, are coupled
at all pairs and triples of times. This can be treated to some
extent like a many fingered time, as we will now see.

The canonical commutation relations for the $B$'s have the
structure,  for $I<J$ and $K<L$,
\f
[  B_{I \alpha}^{J \beta } (x_{I}, t, x_{J}, s)_{AB} , 
\Pi_{K}^{\ L}(x_{K},t;x_{L},u)_{\gamma}^{\delta CD} ]
=\delta^{K}_{I}\delta^{J}_{L}\delta(s,u)
\delta^{2}(x_{K},x_{I}) \delta^{2}(x_{J},x_{L})
\delta_{\alpha}^{\delta}\delta^{\beta}_{\gamma}\delta_{AB}^{CD}
\label{B[]}
\ff

To realize these quantum mechanically we are going to have
to have a Hilbert space for every set of $P$ times
$t_{I}$.  We will call this space $H(t_{I})$.  For
each set of $t_{I}$ we will have operators that act on this
space that satisfy (\ref{B[]}).  This kind of structure is
common to certain histories formulations of quantum theory,
such as those studied in \cite{hpo,FM4,FM5}.  In those papers
it is shown that this kind of multitime quantum theory is
natural for histories formulations of
systems with spacetime diffeomorphism invariance.

The form of the Poisson brackets suggests that we quantize in a 
generalization of
the connection representation in which the states are of the form 
\f
\Psi[\{ t_{I}\}]=\Psi [ A_{I}(t_{I}), B_{I}^{J}(t_{I},t_{J})]
\label{states1}
\ff
for $J>I$.  

The first step in the quantization will be to find solutions to the
Gauss's law constraint (\ref{gauss}). 
To define this we note that the states are defined 
to be functions only of the configuration variables at
a set of $P$ fixed times $t_{I}$.  Thus, 
when acting on a state $\Psi (t_{I})$ the integral
$\int ds$ in \ref{current} will for each $J$ only
pick up times in that list.  The $\delta (s,t)$ in
(\ref{B[]}) absorbs the integral over $s$; thus, the action
of $\hat{j}^{I}(x_{I})$ on states has the form
\begin{eqnarray}
\hat{j}^{I}(x_{I},t_{I})  &&\Psi [ A^{I}(t_{I}) , B_{I}^{J}( t_{I},t_{J}) ]  
\\
&& \equiv 
\left \{ \sum_{J > I}\int  d^{2}x_{J}  
 B_{I}^{J  } (x_{I}, t_{I}, x_{J}, t_{J})
  {\delta \over \delta B_{I }^{J  }(x_{I}, t_{I}, x_{J}, 
  t_{J})}
  \right \}
\Psi [ A^{I}(t_{I}) , B_{I}^{J}( t_{I},t_{J}) ]   \nonumber
\label{quantgauss2}
\end{eqnarray}

A natural set of solutions to these equations
may be obtained by making an ansatz
\f
\Psi [ A_{I}(t_{I}), 
B_{I}^{J}(t_{I},t_{J})]=
\chi [B_{I}^{J}]  
 \prod_{K}\Phi_{K}[A^{K}  ] 
\label{ansatz1}
\ff
To solve this we want to find the action of the operator representing
the current (\ref{current}), which acts as
\begin{eqnarray}
&&\hat{j}^{K P}_{Q}(x_{K}) \chi [B_{I}^{J}]
 \\
&&= \left \{  
\sum_{I<K}\int d^{3}X_{I}Tr_{Sp(2)} B_{IQ}^{KR} \cdot 
{\delta \over \delta B_{IR}^{KP}}
- \sum_{K<J}\int d^{3}X_{J}Tr_{Sp(2)}\cdot B_{KQ}^{JR}
{\delta \over \delta  B_{KR}^{JP}  }
\right \} \chi [B_{I}^{J}]  \nonumber.
\end{eqnarray}

It is clear that there are solutions to the Gauss law which involve
finite products of $B_{I}^{J}$'s. In such a state, for each 
pair of tori $T^{2}_{I}$ and $T^{2}_{J}$,  there
will be a finite number of points 
$(p_{I}^{a},p_{J}^{a}) \in T^{2}_{I} \times T^{2}_{J}$ for
$a=1,\ldots, n_{IJ}$ on which there
are $B_{I}^{J}$'s in the product (\ref{ansatz1}). We will
call these punctures.
The current will then have the form
\f
{\cal J}^{K P}_{Q}(x_{K})= \sum_{I <K }\sum_{a=1}^{n_{IK}}
\delta^{2}(p_{K}^{a}, x_{K}) {\cal J}^{a} [B_{I}^{K}] + 
\sum_{K<J }\sum_{b=1}^{n_{KJ}}
\delta^{2}(p_{K}^{b}, x_{K}) {\cal J}^{b}[B_{K}^{J}]
\ff
where the currents ${\cal J}^{a} [B_{I}^{K}]$ depend on the
$B$'s as indicated.
As a consequence  $\Phi_{K}(A^{K})$ satisfies the condition
\f
{k \over 2\pi} \hat{F}_{12}^{K}(x_{K})\Phi_{K} (A^{K}) =
\left (  \sum_{I <K }\sum_{a=1}^{n_{IK}}
\delta^{2}(p_{K}^{a}, x_{K}) {\cal J}^{a} [B_{I}^{K}] - 
\sum_{K<J }\sum_{b=1}^{n_{KJ}}
\delta^{2}(p_{K}^{b}, x_{K}) {\cal J}^{b}[B_{K}^{J}] \right ) 
 \Phi_{K}(A^{K})
\ff
But this is a familiar equation from quantum Chern-Simons
theory \cite{ed-cs,ms,louis2d3d,verlinde}. The connection on each tori is flat except
for a finite set of punctures at which there is a delta function
contribution, which depends on the $B$'s.  
We can solve this in the standard way, expressing the solutions 
in terms
of conformal blocks or intertwiners of $G$ on the 
punctured two torus.  The dependence of the states on the 
$B$'s will be expressed in terms of the representations labeling
the punctures.
As a consequence the punctures will satisfy braid statistics.
This means care must be taken if we create two punctures on top of each other.
A simple solution to this problem which we will
employ is to construct the states  with all punctures 
distinct and use the recoupling relations of the quantum group 
associated to $SU(16)$  to 
extend their values to the cases of multiple punctures at the same 
point. 

But once the punctures are distinct we can use the gauge symmetry in
(\ref{gauge16}) to reduce the number of independent components of the
$B_{I}^{J}$'s.  Taking $U(i)$ in (\ref{gauge16}) to be valued in
$Sp(2)$ we have transformations,
\f
A_{a}^{I}(x_{I})\tau^{a} \rightarrow U_{I}^{-1}(x_{I}) \cdot A_{a}^{I}(x_{I})\tau^{a}
\cdot U_{I}(x_{I})
\ff
\f
B_{I}^{J}(x_{I},x_{J})  \rightarrow U_{I}^{-1}(x_{I}) \cdot  B_{I}^{J}(x_{I},x_{J})
\cdot U_{J}(x_{J})
\ff
Since the punctures are distinct we can use this gauge freedom to 
diagonalize the $ B_{I}^{J}$.
Given the definition of the momenta,
this gives us, for $I<J$
\f
B_{I}^{J} = 
\left [
\begin{array}{cc}
    b_{+I}^{J} &0  \\
    0 & b_{-I}^{J}  
\end{array}
\right ]
\ff
while  for $J>I$ we define momenta\footnote{It is convenient here to 
take for the time variable the coefficient of $\tau^{2}$ in the 
parameterization we used previously.},
\f
B_{J}^{I} = 
\left [
\begin{array}{cc}
    p_{+I}^{J} &0  \\
    0 & p_{-I}^{J}  
\end{array}
\right ]
\ff
The commutation relations are then. 
\f
[  b_{\pm I }^{J } (x_{I}, t, x_{J}, s)_{PQ} , 
p_{\pm^{\prime }K}^{\ L}(x_{K},t;x_{L},u)^{ RS} ]
=\pm \imath \delta^{K}_{I}\delta^{J}_{L}\delta_{\pm,\pm^{\prime}} \delta (s,u)
\delta^{2}(x_{K},x_{I}) \delta^{2}(x_{J},x_{L})
\delta_{PQ}^{RS}
\label{b[]}
\ff
The two degrees of freedom of the $B_{I}^{J}$  correspond to the
two cases in which the current flows from $I$ to $J$ (with $I<J$) 
in a positive or 
a negative sense. The current is now,
\f
j^{K\ P}_{Q}(x_{K},t_{K})= \sum_{\pm} (\pm 1) \left \{ 
\sum_{I<K}\int d^{3}x_{I}p_{\pm IR}^{KP}  b_{\pm IQ}^{KR}-
\sum_{K<J}\int d^{3}x_{J} p_{\pm KR}^{JP}\cdot b_{\pm KQ}^{JR} 
\right \}
\label{current2}
\ff

We then work with the reduced Hilbert space 
\f
\Psi [ A^{I}_{1}(x_{I},t_{I}) , b_{\pm I}^{\ J} (x_{I},t_{I}; x_{J},t_{J} ) ]
\ff
for all ordered pairs $J>I$.  On this space we will have the kinematical
operators,
\f
\hat{A}_{1}^{I}(x_{I},t_{I}) \Psi = A_{1}^{I}(x_{I},t_{I}) \Psi
\ff
\f
\hat{A}_{2}^{I}(x_{I},t_{I}) \Psi = 
\imath \hbar  {\delta \over \delta A_{1}^{I}(x_{I},t_{I}) }\Psi
\ff
\f
\hat{b}_{\pm I}^{J} (x_{I},t_{I}; x_{J},t_{J} )\Psi =
b_{\pm I}^{J} (x_{I},t_{I}; x_{J},t_{J}) \Psi
\ff
\f
\hat{p}_{\pm I}^{J} (x_{I},t_{I}; x_{J},t_{J}) \Psi = \pm \imath \hbar
{\delta \over \delta b_{\pm I}^{J} (x_{I},t_{I}; x_{J},t_{J})} \Psi
\ff

We can then describe the solutions to the Gauss's law constraint as 
follows.  

\begin{figure}
	\centerline{\mbox{\epsfig{file=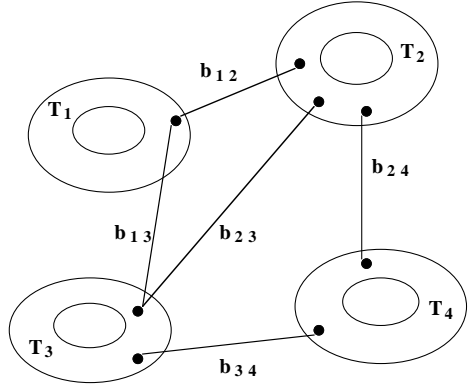,height=2in}}}
	\caption{A typical solution to the Gauss's law constraint
	contains $P$ punctured tori joined in pairs.  On each torus
	the state lives in the space of conformal blocks or
	intertwiners defined by putting each puncture in the
	fundamental representation of $SU(16)$ quantum deformed to level $k$.}
	\label{fig1}
\end{figure}

Let us then pick $R$ pairs of punctures, each member of
a pair is a point on a distinct 2-torus.  The pairs 
are labeled by $w=1,\ldots,R$ and each pair is given by a
point on the 2-torus's $I_{w}$ and $J_{w}$ labeled by,
\f
p_{w}= (p^{+}_{w},p^{-}_{w}) \in T^{2}_{I_{w}} \times T^{2}_{J_{w}}
\ff
Each puncture has a polarity $\epsilon (w) =\pm$.
We then consider an ansatz for the wavefunction,
\f
\Psi [ A^{I}  , b_{\pm I}^{J}]=
\prod_{w}b_{\epsilon (w) I_{w}}^{J_{w}}(p^{+}_{w},p^{-}_{w}) 
\prod_{I}\Phi_{I} (A^{I})]
\ff
These satisfy
\f
\hat{j}^{I}(x_{I})\Psi [ A^{I}  , b_{\pm I}^{J}] = \sum_{w} \epsilon (w)
\left ( {\cal J}_{w}
\delta_{I I_{w}}\delta^{2}(x_{I},p_{w}^{+}) - {\cal J}_{w}
\delta_{I J_{w}}\delta^{2}(x_{I},p_{w}^{-})
\right ) \Psi [ A^{I}  , b_{\pm I}^{J}]
\ff
where ${\cal J}_{w} \in SU(16)$.
This means that $\Phi_{I}(A^{I})$ satisfies the condition
\f
{k \over 2\pi} \hat{F}_{12}^{I}(x_{I})\Phi_{I}(A^{I})=
\sum_{w}  \epsilon (w) \left ({\cal J}_{w}
\delta_{I I_{w}}\delta^{2}(x_{I},p_{w}^{+}) - {\cal J}_{w} 
\delta_{I J_{w}}\delta^{2}(x_{I},p_{w}^{-})
\right ) \Phi_{I}(A^{I})
\ff
We can solve this by using the usual methods from Chern-Simons theory
\cite{ed-cs,ms,louis2d3d,verlinde}.  We may picture a typical state 
in terms of punctured tori whose punctures are connected
in pairs as in Figure (\ref{fig1}).

Finally, we may note that 
once we have fixed the $B_{I}^{J}$ to the
diagonal gauge the quadratic terms in the Hamiltonian, $H_{IJ}^{2}$, 
all  vanish.  This suggests
that the role of these terms is in implementing spatial diffeomorphism
invariance on each of the 2-tori.   Since the choice
of representation of the $Sp(2)$ matrices defined which components
of the matrix variables corresponded to the
spatial coordinates of the tori it is not surprising that there
is in this way a coupling between the $Sp(2)$ gauge freedom
and spatial diffeomorphisms.

We may then finally describe the space
of gauge invariant and diffeomorphism 
invariant states as follows.
We can first describe an auxilliary linear space
$H^{P}_{aux}$ constructed from all ways to join $P$ tori
along pairs of punctures.  Given a set of $R$ punctures
labeled by $\sigma = (I_{w},p^{+}_{w}, J_{w},p^{-}_{w})$ we
join our $P$ tori into a set of compact 2-surfaces by joining
them along each pair of punctures in the set. 
For each $w$ we draw a circle around the
puncture $p^{+}_{w}$ on the $I_{w}$'th torus and another
circle around the puncture $p^{-}_{w}$ on the $J_{w}$'th torus.
We then remove the interiors of each circle and join the two
boundaries together.  The resulting single circle $c_{w}$ is oriented
and labeled with an $s$ , corresponding to a
projection operator which restricts the current flowing accross
$c_{w}$ to be in the spinor representation.  

Once all these operations are completed we have a (generally
disjoint) 2-surface we may call ${\cal S}_{\sigma}$.  The
result of our construction is a Hilbert space
${\cal V}^{P}_{\sigma}$ which is the space of intertwiners
on ${\cal S}_{\sigma}$ subject to the condition that there are
projection operators for the spinor representation on the
$R$ oriented circles $c_{w}$.

For each $P$ and $R$ the space of states lives in the
auxillary Hilbert space given by 
\f
{\cal H}^{P,R}_{aux}=\sum_{\sigma} {\cal V}^{P}_{\sigma}
\ff
where the sum is over all sets $\sigma$ of $R$ pairs of
punctures.  For each $P$ we may let $R \rightarrow \infty$
to define
\f
{\cal H}^{P}_{aux}=\lim_{R \rightarrow \infty} {\cal H}^{P,R}_{aux}¥ .
\label{Haux}
\ff

The auxilliary space (\ref{Haux}) is larger than the physical state 
space in that a given conformal block may be created many different
ways by joining states on punctured tori in this way. As a result 
there is a non-trivial action of  the group, $\cal R$ of modular 
transformations\cite{ms,louis2d3d,verlinde}, or recoupling 
relations\cite{lou} 
of the conformal blocks
on ${\cal H}^{P}_{aux}$.  (There are also linear relations between 
states in different ${\cal H}^{P}_{aux}$ for different $P$.)  We 
then define the physical state space to be,
\f
{\cal H}^{P}_{phys}={\cal H}^{P}_{aux} /  {\cal R}  .
\label{hphys}
\ff
It is not difficult to argue that any conformal block on any
compact surface of genus higher than $1$ can be built up in this
way.  As a result we have
\f
{\cal H}^{P}_{phys}=\sum_{g \geq 1}{\cal V}_{g}
\ff
where ${\cal V}_{g}$ is the space of intertiners on the compact
surface of genus $g$.
For each $P$ this is the physical Hilbert
space of the theory.  We note that becaus there is no limit on the
numbers of punctures all the Hilbert spaces are identical 
for $P\geq 2$.

\begin{figure}
	\centerline{\mbox{\epsfig{file=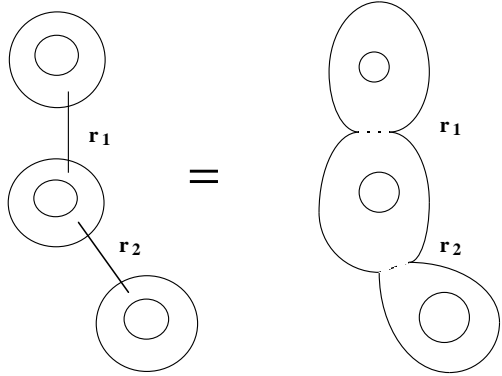,height=2in}}}
	\caption{The translation from states on linked punctured
	tori to states on compact two surfaces.}
	\label{fig2}
\end{figure}

The translation from states on punctured tori to states on
compact surfaces is illustrated in Figure (\ref{fig2}).
We note that each ${ \cal S}_{\sigma}$ can be decomposed
non-uniquely into a set of $P$ 2-tori joined on pairs of circles.
Given a basis for each of the spaces of intertwiners on the
punctured two tori then
each such decomposition gives a basis for ${\cal H}^{P}_{aux}$.  We
arbitrarily fix one such basis, which we will call the reference
basis.  This is necessary because the construction of the Hilbert
space from a set of solutions to the quantum Gauss law constraint produces
such a basis. As we will see Hamiltonian  comes expressed in that
basis.

Finally we note that the state space ( described in \ref{hphys})
is almost the same as that discussed in 
\cite{tubes,pqtubes,mpaper}  The main difference is that the zero
genus state is excluded from the present theory, and in the
proposal of \cite{tubes,pqtubes,mpaper} we considered only
states arising from connected two-surfaces.  This means that the
state space we have arrived at is a natural background
independent extension of the spin network states of quantum
general relativity and supergravity, with 
$SU(16)$ in place of $SU(2)$ or $SU(2|1)$.  

\section{Action of the Hamiltonian and evolution rules}

We now construct the action of the remaining cubic
term (\ref{H3}) in the Hamiltonian on the
Hilbert space (\ref{hphys}).  The hamiltonian is constructed
in terms of the operators $b_{\pm I}^{J}$ and $p_{\pm I}^{J}$
which means their action is defined on the auxillary space
(\ref{Haux}).  The action on the physical space (\ref{hphys}) is
defined by imposing equivalence under modular transformations.
We have
\begin{eqnarray}
\sum_{I<J<K}&\int& dt_{I} dt_{J}dt_{K}H^{IJK}(t_{I},t_{J},t_{K})
\Psi (\{ t_{I}\} ) =  \nonumber \\ 
&& 3 \sum_{\pm}  \int d^{2}x_{I}d^{2}x_{J}d^{2}x_{K}  Tr_{G} 
\left \{  \sum_{I<J<K}  b_{\pm I}^{\ J} b_{\pm J}^{\ K} p_{\pm I}^{\ K}
+ { 4\pi \over k} \sum_{I<K<J}  b_{\pm I}^{\ J} p_{\pm K}^{\ J} p_{\pm I}^{\ K}
\right \}
\label{H!!!}
\end{eqnarray}

\begin{figure}
	\centerline{\mbox{\epsfig{file=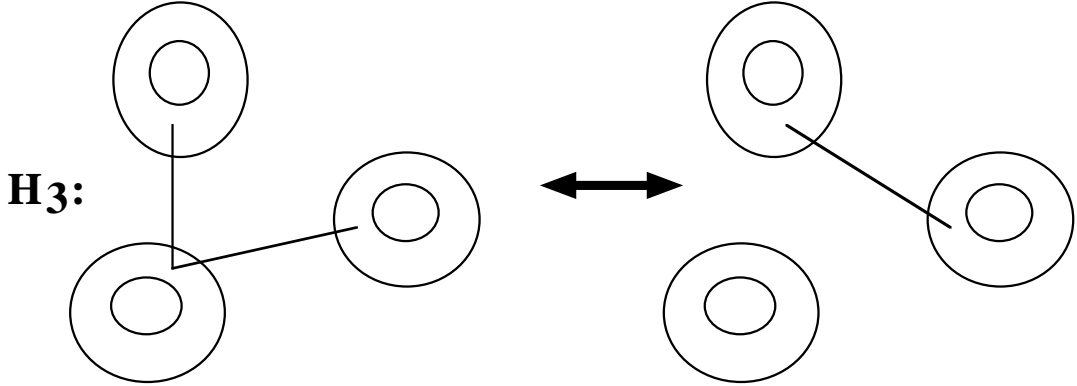,height=2in}}}
	\caption{The action of the cubic term in the Hamiltonian (\ref{H!!!}).}
	\label{fig3}
\end{figure}

The action of $H_{3}$ on states 
is given in Figure (\ref{fig3}).
$H_{3}$ has two kinds of actions. Two of the terms eliminate two links
labeled by the $b$'s and
creates a new one as shown in Fig. (\ref{fig3}).  In this case 
the action of $H$ eliminates a positive and negative puncture on
a single torus. It then induces a map from 
${\cal V}_{s\otimes \bar{s} , r_{2},\ldots}\rightarrow
{\cal V}_{r_{2},\ldots}$, where $s$ is the fundamental
representation, which is defined by taking the
intertwiner $s \otimes \bar{s}\rightarrow Id$. The
other terms act in the opposite way to  eliminate a
$b_{I}^{K}$ joining tori $I$ and $K$ and create, for every
distinct torus $I \neq J \neq K$ a link from $I$ to $J$ and 
a link from $J$ to $K$. The action on the space of intertwiners
on $J$ then acts in the opposite way, through the
channel $ Id \rightarrow s \otimes \bar{s}$.

\begin{figure}
	\centerline{\mbox{\epsfig{file=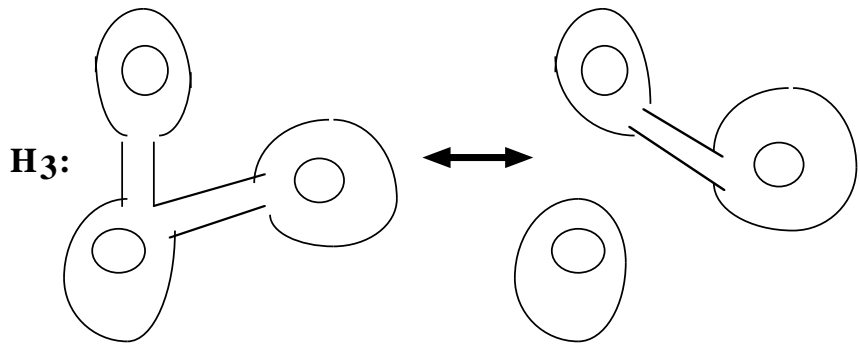,height=2in}}}
	\caption{The action of the cubic term in the Hamiltonian
	translated to states on compact two surfaces.}
	\label{fig4}
\end{figure}

Translated to an action on states on compact two surfaces
rather than punctured tori the action of $H_{3}$ is
illustrated in Figure (\ref{fig4}).

\begin{figure}
	\centerline{\mbox{\epsfig{file=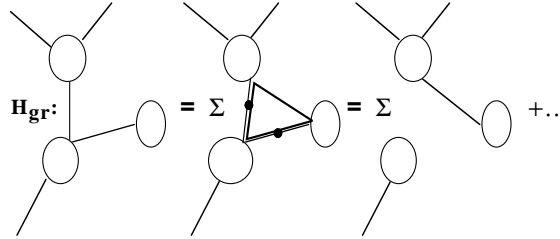,height=2in}}}
	\caption{The action of the Hamiltonian
	constraint on spin network states in quantum
	general relativity indicated schematically.  The
	dependence on representations and the coefficients
	of the different terms are not indicated but can be
	found in \cite{ham1,thomas1}.  The action of the constraint
	induces Wilson lines as shown, inserted into the edges
	as shown by the dots.  The summation indicates a summation
	over certain representations, which are suppressed. The
	$k\rightarrow \infty$ limit has been taken so that the
	compact two surfaces dressed with intertwiners become
	graphs dressed by representations.}
	\label{fig5}
\end{figure}

We may note that the action has exactly the form of the
Hamiltonian constraint for quantum general relativity,
found in \cite{ham1}\footnote{different variants are
discussed in \cite{thomas1}.} with the replacement of
$SU(2)$ by the quantum deformation of $SU(16)$.  
This can be seen in Figure (\ref{fig5}) where we
indicate schematically the action of the Hamiltonian
constraint of general relativity found in \cite{ham1,thomas1}.

This means that if we restrict $SU(16)$
to an $SU(2)$ subgroup, the dynamics predicted by the
cubic matrix model will be of the same form as 
that found from canonical quantization
of quantum general relativity.  In terms of the dual
picture of \cite{FM2} this gives the $1\rightarrow 3$
and $3 \rightarrow 1$ Pachner moves.  
Both the $1\rightarrow 3$ and $3 \rightarrow 1$ moves
arise in the theory with the same amplitude, showing
that the Hamiltonian is hermitian.

\begin{figure}
	\centerline{\mbox{\epsfig{file=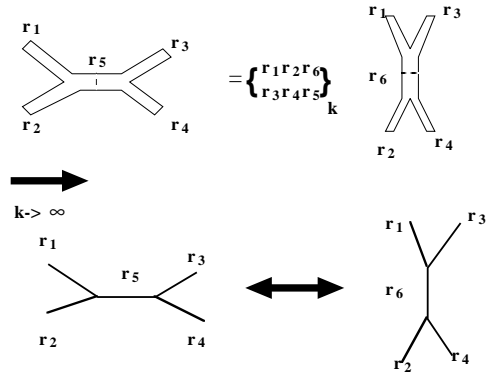,height=2in}}}
	\caption{The change of basis formula for finite
	$k$ induces new $2\rightarrow 2$ moves in the limit
	$k\rightarrow \infty$.}
	\label{fig6}
\end{figure}

The fact that the present theory is at finite $k$ means,
moreover, that the difficulty found for quantum general relativity,
in which there is no long range propagation of information,
and hence no chance of a classical limit\cite{trouble1,trouble2}
need not be present.  The reason, as pointed out in 
\cite{tubes}
is that at finite $k$ the missing moves which are required
to have propagation of information are the $2\rightarrow 2$
Pachner moves, shown in Figure (\ref{fig6}).  However, these moves
are present for all theories with finite $k$ as they
are just changes of basis in the space of intertwiners
as illustrated in Figure (\ref{fig6}).  Thus, if we consider
a sequence of moves for finite $k$ induced by $H_{3}$ we may
insert at any point the change of basis shown in Fig. 5.  For
finite $k$ this is just a change of basis and the history in
terms of finite $k$ states is unchanged whether we do it or not.
But this operation does not commute with the limit $k\rightarrow 
\infty$. This means that the histories that must be included in
the sum over histories at
$k\rightarrow \infty$ on the spin network states includes the
insertion of all possible finite $k$ change of basis moves. Since
these are no longer changes of bases for $k=\infty$ this is
equivalent to the insertion in the $k=\infty$ hamiltonian
constraint of the change of basis move, with the amplitudes
given by the $k\rightarrow \infty$ limit of the $6j$ symbol.
This is equivalent to saying that the theory has these
$2\rightarrow 2$ moves with the amplitudes given by the $BF$
or Crane-Yetter-Ooguri theory, while the $1\rightarrow 3$
and $3\rightarrow 1$ moves have the non-topological form given
by $H_{3}$.
However, with the inclusion of these $2\rightarrow 2$ moves the
problem with lack of long range correlations found in 
\cite{trouble1,trouble2} is solved, as pointed out in 
\cite{CM1,FM2}.

Finally, we note that the fact that the $2\rightarrow 2$ moves have
a different amplitude than the $1\leftrightarrow 3$ move means that
there is a dependence of the amplitudes on the causal structure of the
quantum history\cite{FM2,tubes}.  The theory does not have the crossing
symmetry required by a Euclidean quantum gravity 
theory\cite{CM1,carlofoam}, it is then intrinsically a causal theory
in the sense of \cite{FM2,tubes}.

\section{Compactification to the dWHN-BFSS model}

In this section we describe very briefly a different
compactifcations which appears to 
lead to the dWHN-BFSS model at the one loop level.  This will
be discussed in detail in \cite{1loop}.
The required compactification involves a simultaneous compactification
on three directions, but it is different than the one described
in section 4.  There we considered a compactification that
broke the $U(P)$ symmetry involving the indices $I,J,\ldots$
while preserving the whole $SU(16)$ symmetry.  Now we consider
a compactification that does the reverse: we preserve the
$U(P)$ symmetry, but break the $SU(16)$ symmetry.

To do this we break the fields in terms of $SU(16)$ traces
and trace-free parts:
\f
A_{aI i_{0}i_{1}i_{2}P}^{\ J j_{0}j_{1}j_{2}Q}= 
\delta^{Q}_{P}{\cal A}_{aI i_{0}i_{1}i_{2}}^{\ J j_{0}j_{1}j_{2}}
+ \tilde{A}_{aI i_{0}i_{1}i_{2}P}^{\ J j_{0}j_{1}j_{2}Q}
\ff
where $\tilde{A}_{aI i_{0}i_{1}i_{2}P}^{\ J j_{0}j_{1}j_{2}P}=0$.  

We begin as before by expanding around the classical solution
\f
{\cal A}_{aI i_{0}i_{1}i_{2}}^{\ J j_{0}j_{1}j_{2}}= \delta_{I}^{J}
({\partial}_{a})_{ i_{0}i_{1}i_{2}}^{\  j_{0}j_{1}j_{2}} ; \ \ 
\tilde{A}_{aI i_{0}i_{1}i_{2}P}^{\ J j_{0}j_{1}j_{2}Q}=0
\ff
However, what is different in this case is that we define
the compactification in such a way that all 
of the $SU(16)$ components
of the fields becoming $SU(P)$ matrices. To do this we employ a 
parameterization that preserves the $U(P)$ gauge symmetry
which is
\f
A_{aI i_{0}i_{1}i_{2}P}^{\ J j_{0}j_{1}j_{2}Q}= 
\delta^{Q}_{P}\left ( \delta_{I}^{J}
{\partial}_{a i_{0}i_{1}i_{2}}^{\  j_{0}j_{1}j_{2}}
+a_{aI i_{0}i_{1}i_{2}}^{\ J j_{0}j_{1}j_{2}} \right )
+ \tilde{A}_{aI i_{0}i_{1}i_{2}P}^{\ J j_{0}j_{1}j_{2}Q}
\ff

We will also restrict the fields to real values, so that the
symmetry is reduced from $SU(16,16|1)$ to $Osp(1|32)$.

We now compactify on the three directions associated with $x^{+}$,
$x^{-}$ and $x^{2}$. This is similar to the compactification we
used up till this point, but we will treat it in a rather different
manner.  Rather than considering the limit in which all three 
compactification radii go to infinity, which leads us to a Chern-Simons
type theory, we will keep the compactification radii
$L_{2,-}=(2M_{2,-}+1) l_{Pl}$ small, while taking only the third radius
$L_{+}=(2M_{+}+1) l_{Pl}$ large.  This breaks the $SO(2,1) =Sp(2)$ 
symmetry.  
The gauged action can be written as
    \begin{eqnarray}
I^{gauged}&=& {k \over 4\pi } \int_{T^{3}}
dx_{+}dx_{-}dx_{2}Tr_{U(P)} \left \{ Y^{CS}(a) \right. \nonumber \\
&&+\epsilon^{abc}\left ( 3 
\tilde{A}_{aP}^{\ Q}{\cal D}_{b}\tilde{A}_{cQ}^{\ R}
+ \tilde{A}_{aP}^{\ Q}\tilde{A}_{bQ}^{\ R}\tilde{A}_{cR}^{\ P} 
\right ) 
\nonumber \\
&&+Y^{3} 
+\eta^{ab} \tilde{Y}_{P}^{Q}( \{ {\cal D}_{a}, \tilde{A}_{bQ}^{P} \}
+ \tilde{A}_{aQ}^{R}\tilde{A}_{bR}^{P})
\nonumber \\
&&+ \left.  \chi^{\dagger P}\tau^{a}_{PQ}{\cal D}_{a} \chi^{Q}
\right \} \nonumber
\end{eqnarray}
although we must remember that as we have not taken the 
compactification radii to infinity what is really meant is the
cutoff version of this theory with $2M_{a}+1$ modes in each
direction.  An important consequence of not taking the limit
of infinite compactification radii is that the theory
knows about the $2+1$ dimensional metric $\eta^{ab}$.

The next step is to compute the effective action.  This will
be described in \cite{1loop}, here we only  discuss how the
symmetries of the theory  constrain its form.

The form of the effective potential will be governed by the 
symmetries of the action which are left unbroken by the
compactification.  The local  bosonic symmetries
includes $U(P)$.  In the  case
of compactifications with independent compactification radii
$L_{+}\neq L_{-}\neq L_{2}$, the $SU(16,16)$ symmetry has been
broken down into a subgroup, which is $SO(9)$.  This can be
seen from the fact that in a standard parameterization of the
components of the matrix $M$ in terms of $32$ component
$\Gamma$ matrices for $Spin(10,1)$, the $+,-$ and $2$ components that
defined the three $\tau^{a}$ matrices that generate the
$Sp(2)$ symmetry are in the $0$, $10$
and $0 \wedge 10$ boost directions \cite{cubic}.  

There are in addition the remaining
translation symmetries (\ref{translations}) involving the tracefree parts,
$\tilde{A}_{aP}^{\ Q}$.  These are symmetries of the action in 
the case that we restrict all fields to the real slice.   This is
one reason we have restricted the fields to the real case.  Finally,
one can check that $16$ of the $32$ supersymmetries have been
broken, so that there remain $16$ unbroken supersymmetry
generators.

The lowest dimensional terms that can appear in the effective
action, consistent with these symmetries are,
\begin{eqnarray}
I^{1 \ loop}&=&  \int_{T^{3}}
dx_{+}dx_{-}dx_{2}Tr_{U(P)} \left \{
f_{ab}f_{cd}\eta^{ac}\eta^{bd}+ 
[{\cal D}_{a},\tilde{A}_{b}] [{\cal D}_{c},\tilde{A}_{d}]\eta^{ac}\eta^{bd}
\right.  \nonumber \\
&& \left. + {(\cal D}_{a}\tilde{A}_{b}\eta^{ab})^{2} 
 + fermions + interactions \right \} 
\nonumber
\end{eqnarray}

We now consider the limit
\f
L_{-},L_{2}\rightarrow l_{Pl}, \ \  L_{+}\rightarrow \infty .
\ff
At the same time we boost to the infinite momentum frame in the
$+$ direction.  All fields decouple except those that have the
maximal number of $\partial_{+}$ derivatives.  These turn out to
be the $\tilde{A}_{-}$ and $\Psi$ fields.   

In this limit the only terms that survive in the kinetic energy
are
\f
I^{1 \ loop}= \int dx_{+}Tr_{U(P)} \left \{
(\partial_{+}\tilde{A}_{-})^{2} + \Psi^{P \dagger}\partial_{+}\Psi_{P} 
\right \} 
\ff
We see that the fields which remain consist of one
$16$ component fermion together with the symmetric
tensor $A_{-}^{PQ}$.   
This decomposes as
\f
A_{-PQ}=\Gamma_{PQ}^{\mu}X_{\mu}+ \Gamma_{PQ}^{\mu\nu \rho \sigma} 
X_{\mu \nu \rho \sigma}
\ff

The $X_{\mu}$ are the nine transverse matrices of the
dWHN-BFSS theory, which correspond to the $D0$ brane coordinates
as well as to the light cone gauge coordinates of the embedding
of a membrane. The $X_{\mu \nu \rho \sigma}$ are additional degrees of
freedom that do not appear in that model.  

The leading terms in a derivative expansion of the effective
action will be  completed by interaction
terms amongst these degrees of freedom.  These will be determined
by the unbroken symmetries, which are precisely the symmetries of
the dWHN-BFSS theory.   The action involving $\Psi^{P}$
and $X^{\mu}$ with these symmetries is exactly the 
dWHN-BFSS theory.  One can check that this is extended
by terms involving the four form field $X_{\mu \nu \rho \sigma}$.
The full action invariant under the translations (\ref{translations}),
 $SO(9)$ rotations and $16$ supersymmetry generators is
\begin{eqnarray}§
I^{1 \ loop}&=& \int dx_{+}Tr_{U(P)} \left \{
(\partial_{+}\tilde{A}_{-})^{2} + \Psi^{P \dagger}\partial_{+}\Psi_{P}
\right. \nonumber \\
&& \left. + \Psi^{P \dagger}[\tilde{A}_{- PQ}, \Psi^{Q} ] 
+ 
\tilde{A}_{-[P}^{\ Q}\tilde{A}_{-R]Q}\tilde{A}^{PS}_{-}\tilde{A}^{R}_{-S}
\right \}  \\ \nonumber
&=& \int dx_{+}Tr_{U(P)} \left \{
(\partial_{+}X_{\mu})^{2} + (\partial_{+}X_{\mu \nu \rho \sigma})^{2}
+ \Psi^{P \dagger}\partial_{+}\Psi_{P}
\right. \nonumber \\
&&  + \Psi^{P \dagger} \Gamma^{\mu PQ}[X_{\mu}, \Psi^{Q} ] 
+  \Psi^{P \dagger} \Gamma^{\mu \nu \rho \sigma PQ}[X_{\mu \nu \rho \sigma}, 
\Psi^{Q} ]  
\nonumber \\
&&+ [X^{\mu}, X^{\nu}][X_{\mu}, X_{\nu}]
\nonumber \\
&&+
\left. [X^{\mu \nu \rho \sigma}, X^{\alpha \beta \gamma \delta}]
[X_{\mu \nu \rho \sigma}, X_{\alpha \beta \gamma \delta}] +
[X^{\mu}, X^{\alpha \beta \gamma \delta}][X_{\mu}, 
X_{\alpha \beta \gamma \delta}]
\right \}  
\end{eqnarray}

One can check that the  $X_{\alpha \beta \gamma \delta}$
terms do not break the supersymmetry of the dWHN-BFSS theory.
Instead,the symmetry algebra is extended by central charges
that exist for any $P$.  In the standard dWHN-BFSS model
these central charges appear only in the limit
$P \rightarrow \infty$\cite{infinitebranes}.  In that case the central
charges are interpreted to
describe certain components the $5$-brane.  This suggests
that the new degrees of freedom $X_{\alpha \beta \gamma \delta}$
provide an additional description of the $5$-brane, wrapped on
the $X^{+}$ and $X^{-}$ directions.  This will be discussed
in more detail elsewhere.

\section{The duality between strings and loops}

From the results of the last several sections it is possible
to describe precisely how the BFSS matrix model and the 
$SU(16)_{q}$ extension of loop quantum
gravity give equivalent descriptions of the same degrees of freedom.
We can do this by tracing how each description is derived
from a compactification of the cubic matrix model. 

In both cases we make use of the decomposition
$SU(16,16|1) \rightarrow Sp(2) \oplus SU(16|1)$ and 
decompose the fields in terms of the basic degrees of freedom,
\f
A_{a I i_{a} P }^{\ J j_{b} Q}
\ff
The $a$ is the $Sp(2)$ index and the $P$ and $Q$ are
$SU(16)$ indices.  We have then decomposed the  $N\times N$
components of the matrices in terms of two sets of indices
so that
\f
i = (i_{a},I)
\ff
where $i_{a}=1,\ldots, M_{a}$ are the indices that will give
rise to compactification on a three torus, and $I=1,\ldots, P$ defines
a remaining $U(P)$ symmetry. 

In both cases we begin by defining a 
compactification
on a $3$-torus, leaving aside a  $U(P)$ symmetry.   Each 
compactification 
is then defined by an expansion around the classical solution
\f
(A^{0}_{a})_{ I i_{a} P }^{\ J j_{b} Q}
= \delta_{I}^{J}\delta_{P}^{Q}(\partial_{a})_{i_{a}}^{j_{a}}
\label{primordial}
\ff
Each then defines a fluctuating field
\f
A_{a I i_{a} P }^{\ J j_{b} Q} =
\delta_{I}^{J}\delta_{P}^{Q}(\partial_{a})_{i_{a}}^{j_{a}} +
 {\cal A}_{a I i_{a} P }^{\ J j_{b} Q}
\ff
The stringy and loopy descriptions then part company.  In each case one 
of the two remaining symmetries, which are $U(P)$ and $SU(16)$, are
broken and the other is  treated as a gauge symmetry in the background
created by the compactification.  The two 
compactifications differ as to 
which components of the fluctuating field 
${\cal A}_{a I i_{a} P }^{\ J j_{b} Q}$ are treated as a gauge field
and which are taken as a matter-like field.  

There is freedom to chose which symmetry appears as an ordinary
gauge symmetry because 
before compactification the fields
have three index types: $A_{a i_{a}IP}^{j_{a}JQ}$.
The $i_{a}$ describe the fourier modes of the spatial
dependence on the $3$ torus, while the $I$ and $P$ parameterize
respectively $SU(P)$ and $SU(16)$.  The gauge symmetry as it
appears as a local symmetry in the three dimensional field theory
on the $T^{3}$'s comes from a symmetry of the form of either
(\ref{gaugeN}) or (\ref{gauge16}).  But there are now three
sets of indices, which must be grouped into two sets. The
first set determines which indices the gauge parameters depend
on, while the second determines the gauge group.  For example
if we group the $i_{a}$ with the $I$ then we have gauge
transformations of the form
\f
 A_{i_{a}I}^{\ j_{a}J} \rightarrow U(i_{a}I) \cdot A_{i_{a}I}^{\ j_{a}J} 
\cdot U(j_{a}J)^{\dagger}
\label{gauge16again}
\ff
where $U(i_{a}I) \in SU(16)$.  This gives an $SU(16)$ gauge
symmetry that acts locally in each of the tori labeled by $I$.
This is the symmetry of the background independent multi-tori
compactifications.  In this representation of the theory
the $U(P)$ symmetry is hidden.  It is possible that it is
expressed by the recoupling relations of the intertwiners
of $SU(16)_{q}$
(equivalently the modular group of the 2-surfaces 
acting on the conformal
blocks).

In this case we defined the components of the fluctuating fields
as
\f
 {\cal A}_{a I i_{a} P }^{\ J j_{b} Q}= \delta_{I}^{J}
  A_{a P i_{a} }^{\ I Q j_{b}} +
   B_{a I i_{a} P }^{\ J j_{b} Q}
\ff
We treated the diagonal components $ A_{a P i_{a} }^{\ I Q j_{b}}$
 as gauge fields for the $SU(16)$ symmetry. There are
$P$ of them, corresponding to the diagonal components of
$U(P)$ matrices, so we have a separate gauge invariance on
each of  $P$ tori.  The off-diagonal (in $U(P)$) components
$B_{a I i_{a} P }^{\ J j_{b} Q}$ defined only for $I\neq J$
define interactions between the $P$ Chern-Simons theories
defined by the diagonal components  $ a_{a P i_{a} }^{\ I Q j_{b}}$.
The $U(P)$ symmetry is broken by the choice of polarization, which
requires an arbitrary ordering of the $P$ Chern-Simons theories.

The other choice is to make manifest the gauge symmetry that comes
from grouping the $i_{a}$ indices with the $SU(16)$ indices.
In this case we write,
\f
 A_{i_{a}P}^{\ j_{a}Q} \rightarrow W(i_{a}P) \cdot  A_{i_{a}P}^{\ j_{a}Q}
\cdot W(j_{a}Q)^{\dagger}
\label{gaugePagain}
\ff
where $W(i_{a}P) \in SU(P)$. In this case we get a local 
$SU(P)$ gauge symmetry on a single torus.  This leads to the case
of the stringy compactification, which is expressed in terms
of a local $SU(P)$ gauge theory, which we argued leads to the
dWHN-BFSS matrix model.

In this case we 
we defined the components of the fluctuating field as
\f
 {\cal A}_{a I i_{a} P }^{\ J j_{b} Q}= \delta_{P}^{Q}
  a_{a I i_{a} }^{\ J j_{b}} +
   \tilde{\cal A}_{a I i_{a} P }^{\ J j_{b} Q}
\ff
where $ \tilde{\cal A}_{a I i_{a} P }^{\ J j_{b} P}=0$.  
The scalar (in $SU(16)$ terms) components, $a_{a I i_{a} }^{\ J j_{b}}$
define the fluctuations in the compactification radii, and then
must become the string theory moduli.  The theory is treated from that
point on as an $U(P)$ gauge theory, where the scalars 
$a_{a I i_{a} }^{\ J j_{b}}$ play the role of gauge fields in 
the compactified directions and the $SU(16|1)$ symmetry is broken
down to the superpoincare algebra in $10+1$ dimensions or the
super-euclidean algebra in $9$ dimensions.  

The different theories are treated differently in other ways,
particularly in the fact that to get the standard matrix
models two or three of the compactified radii must be taken to
the Planck scale, leading to a low energy theory defined in either
$0+1$ or $0+0$ spacetime dimensions (in the dWHN-BFSS and IKKT
cases, respectively.)  But the essential difference between
them is defined by these two different coordinatizations of the
fluctuating fields around the classical solution
(\ref{primordial}).  

As a consequence of the identifications defined here we have
a genuine translation between the loopy and stringy descriptions
of the kinematics and dynamics of the cubic matrix model.  This
may be used to translate problems from one description to the other.
For example, the $P \rightarrow \infty$ limit which is problematic
for the conventional matrix models is clearly related to the
limit of loop quantum gravity in which the universe grows infinitely
large.  On the other side we can say that the continuum limit of
the loopy description may involve a restoration of the $U(P)$ 
symmetry which is broken by the quantization described here.  The
correspondence may also be used to translate the $D$-brane
description of black hole horizons into the loop quantum
gravity description\cite{loopbh}, which is based on describing the
state space on the horizon in terms of conformal blocks\cite{linking}.
This may make possible a description of black holes which is
not restricted to the near extremal case of positive specific heat.
Similarly, we may try to use the correspondence to extend the
description of boundaries with non-zero cosmological constant
given in \cite{linking} to arrive at a detailed description of
the $AdS/CFT$ correspondence in $3+1$ dimensions.

The duality can also be expressed by considering how the fundamental
excitations are described in the two pictures.  In
the multi-Chern-Simons compactifications,  the fundamental
degrees of freedom are the pairs of punctures 
which are created or destroyed by the dynamical terms
in the Hamiltonian. It is then tempting to see them as the 
background independent analogue of $D_{0}$ branes.  There is, in 
fact, some more direct 
evidence that these excitations are related to $D_{0}$ branes, which 
is described in \cite{mpaper}.  We then have two different 
representations of these degrees of freedom, either in terms of the
light cone gauge components of the matrices $A_{- I}^{\ J}$, which
lead to the standard matrix description of $D0$ branes, or
as the operators which create and annihilate punctures which 
join the $P$ tori.
By tracing through the correspondence we have just discussed we can
see explicitly how the two descriptions of the fundamental degrees
of freedom may be translated into each other. 
It is then perhaps fitting to call 
this the string/loop duality.
\begin{figure}
	\centerline{\mbox{\epsfig{file=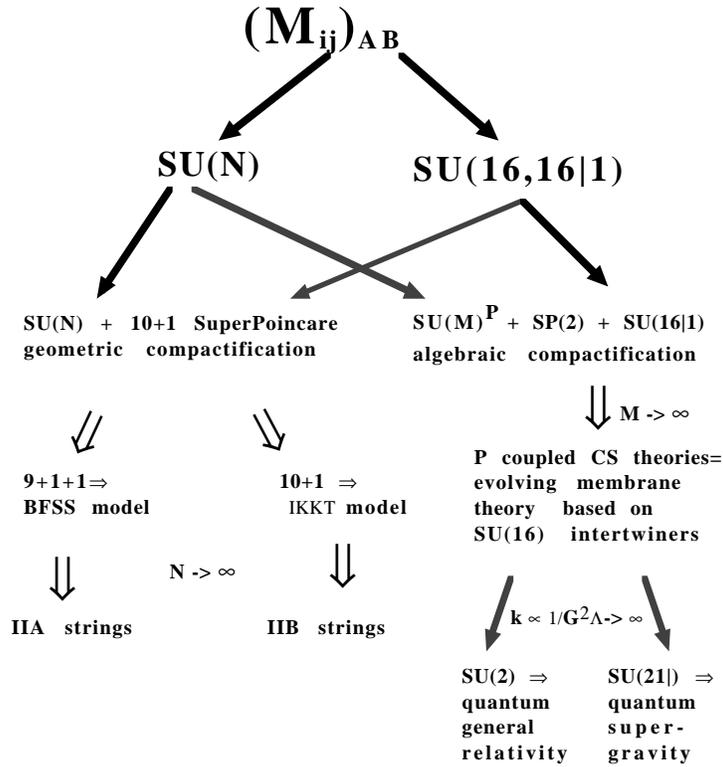,height=4in}}}
	\caption{A string/loop duality, suggested by the results
	of this paper and \cite{cubic}.}
	\label{dualityfig}
\end{figure}

\section{Conclusions}

In this paper we have studied a new kind of compactification for
$\cal M$ theory, which is defined, not in terms of a background
geometry, but in terms of an algebra, that is the fusion
algebra of conformal blocks for the quantum deformation of
$SU(16)$ on arbitrary 2-surfaces.  We may then call this
an algebraic compactification.  It is clear from the construction
that many other algebraic compactifications can be defined 
corresponding to the reduction of the representation theory
of quantum deformed $SU(16)$ to the representation theory
of its subalgebras.

There are many issues raised by this formulation.  Some are
technical.  The most pressing of these include the need to study in detail
the quantum group extension of $SU(16)$ defined by a  
Chern-Simons theory with a connection valued in that algebra
and the need to develop the details of its representation theory.
Another set of issues to be discussed in \cite{superthis} involve the
details of the supersymmetric extension of the results described
here.  We want to understand also the way in which the modular group
acting on compact two surfaces is related to the $U(P)$ symmetry
that was broken in the multi-Chern-Simons compactification.  The
computation of the effective action which we sketched above
will be discussed in detail in \cite{1loop}.  The relationship
between the real and complex forms of the theory needs more 
investigation.
Finally, there remain subtle issues associated with the
role of the $Y$ degrees of freedom.

On the conceptual side, the role of the many fingered time
of the form found here needs further study.  The action
appears to be non-local in time, but the multi-time quantization
discussed here seems to lead to a sensible notion of a history.
One may begin with an initial state at a time $t_{I}=0$ for
all $t_{I}$ and then act repeatedly by the cubic hamiltonian
to evolve, on each action, triples of Hilbert spaces forward
in time. In this way one can generate a history, which  seems to
be of a form which is closely related to that described in \cite{tubes}.
In that case acting repeatedly with the moves generates a causal
history, of the general form studied in \cite{FM4,FM5}.
The causal structure results from the ordering dependence
of the action of the different terms in the Hamiltonian, the
time labels themselves seem irrevelevant apart from ordering.
This structure is very suggestive but deserves further study.
The relationship to the histories projection operator formulation
developed in \cite{hpo} is also very suggestive.

Another interesting issue to discuss is the relationship
of this formulation with the holographic principle and the 
related problem of the interpretation of quantum
theories of cosmology.  The construction in \cite{tubes,pqtubes,mpaper}
was motivated in part by the holographic principle, which we
showed in \cite{linking,hologr} appears naturally in loop
quantum gravity when boundaries are considered.
It is interesting to  consider each one of the Hilbert spaces 
$H(t_{I})$ to be a screen,
in a background independent formulation of the holographic principle
of the kind described in \cite{screens}.  We may then to try to follow
the argument there and  construct the quantum geometry by
defining the area of each screen to be the log of the dimension
of its space of intertwiners, as in \cite{linking}.

When we take the limit $k\rightarrow \infty$ we reproduce,
as argued in \cite{tubes} exactly the spin network states of
quantum general relativity or supergravity.  Moreover, the cubic
action gives rise to evolution rules which are exactly of the
same form as follow from a first principles, canonical quantization
of general relativity\cite{ham1}.  The main difference is that because the
theory is defined at finite $k$, new evolution rules must appear
in the $k\rightarrow \infty$ limit, exactly of the form required
to cure a major problem of loop quantum gravity, which is the
absence of long ranged correlations at zero cosmological 
constant\cite{trouble1,trouble2}.  

It is very interesting to note that there are other derivations of
path integral formulations for loop quantum gravity from matrix
models\cite{carlofoam,CM1,robertonew}.  It is possible that
there is a direct derivation from the cubic matrix model to matrix
models of the form used in those derivations, which is induced by
quantum corrections to the present model. This is presently under
investigation.  Also of interest is the question of whether the
topological field theory parameterization of the degrees of
freedom of $11$ dimensional supergravity introduced in \cite{11tqft}
can be derived directly from the cubic matrix model studied here.

Finally, any approach to a background independent formulation of
$\cal M$ theory must answer the question of how the particular
structures which seem required by string theory for perturbative
consistency of a quantum theory of gravity are picked out at the
more fundamental, background independent level.  It is quite
possible that the theory presented here has a more fundamental
formulation in which the choice of algebra is not arbitrary.  
The  possibility of a reformulation in terms of
Jordan algebras and octonions comes naturally to mind.
The $SU(16,16)$ structure reminds one of an octonionic
extension of twistor theory. There is also an intriguing
similarity to Chern-Simons inspired formulations of string
field theory\cite{sft} that deserves further investigation.

\section*{ACKNOWLEDGEMENTS}

I am  grateful to Louis Crane, Bartomeu Fiol, Willy Fischler, 
Mike Green, Chris Hull, Chris Isham, Yi Ling, 
Fotini Markopoulou,  
George Minic, Mike Reisenberger, Kelle Stelle  and Dennis Sullivan
for discussions, suggestions and encouragement 
during the course of this work. In addition, Richard Levine 
contributed a number of extremely helpful observations and 
corrections.  I am grateful also 
for hospitality at the Institute in Theoretical Physics in Santa 
Barbara, where this work was begun, and to the theory group at Rutgers
University, where it was completed.  This work was supported by 
the NSF through grant PHY95-14240 and by an SPG grant. 
I would also like to thank the 
Jesse Phillips Foundation for support and encouragement.

\end{document}